\documentclass[useAMS,usenatbib]{mn2e}

\usepackage[squaren]{SIunits}
\usepackage{amsmath}
\usepackage{graphicx}
\usepackage{caption}
\usepackage{hyperref}

\title[Dust in three dimensions in the Galactic Plane]{Dust in three dimensions in the Galactic Plane}
\author[R. J. Hanson et al.]{R. J. Hanson$^{1}$\thanks{E-mail:
hanson@mpia.de}, C. A. L. Bailer-Jones$^{1}$\thanks{E-mail:
calj@mpia.de}, W. S. Burgett$^{2}$, K. C. Chambers$^{2}$,
\newauthor K. W. Hodapp$^{2}$, N. Kaiser$^{2}$, J. L. Tonry$^{2}$, R. J. Wainscoat$^{2}$, C. Waters$^{2}$\\
$^{1}$Max-Planck-Institut f\"ur Astronomie, K\"onigstuhl 17, 69117 Heidelberg, Germany\\
$^{2}$Institute for Astronomy, University of Hawaii at Manoa, Honolulu, HI 96822, USA}
\begin{document}

\date{Accepted 2016 September 5. Received 2016 August 23; in original form 2014 November 24}

\pagerange{\pageref{firstpage}--\pageref{lastpage}} \pubyear{2016}

\maketitle

\label{firstpage}

\begin{abstract}
We present three dimensional maps in monochromatic extinction $A_{\rm 0}$ and the extinction parameter $R_0$ within a few degrees of the Galactic plane. These are inferred using photometry from the Pan-STARRS1 and Spitzer Glimpse surveys of nearly $20$ million stars located in the region $l = 0-250^{\circ}$ and from $b = -4.5^{\circ}$ to $b=4.5^{\circ}$. Given the available stellar number density, we use an angular resolution of $\unit{7}{'} \times \unit{7}{'}$ and steps of $\unit{1}{mag}$ in distance modulus. We simultaneously estimate distance modulus and effective temperature $T_{\rm eff}$ alongside the other parameters for stars individually using the method of \citet{Hanson2014} before combining these estimates to a complete map. The full maps are available via the \textit{MNRAS} website.
\end{abstract}


\section{Introduction}
Recently, several new studies analysing the distribution of extinction and dust in the Galaxy have appeared, emphasising the importance of improving our understanding of this key component of the Milky Way Galaxy. Having moved on from the two-dimensional maps that can only characterise the total line of sight extinction \citep[e.g.][]{Schlegel1998}, we can now estimate extinction in three dimensions, utilising several large-scale photometric surveys to infer individual stellar parameters and distances to millions of stars.

\citet{Marshall2006} use red giant stars to map extinction using near infrared data from 2MASS based on a Galactic model. \citet{Gonzalez2011, Gonzalez2012} similarly compare colours of red clump stars to reference measurements in Baade's window to obtain a high-resolution map of the central bulge.

\citet{Berry2012} compare SDSS and 2MASS photometry to the spectral energy distribution from stellar templates, performing a $\chi^2$ fit to the data. Analogously, \citet{Chen2014} analyse XSTPS-GAC, 2MASS and WISE data on the Galactic anti-centre.

In recent years, several new methodological approaches have been introduced, in particular Bayesian ones. \citet{Bailer-Jones2011} uses our understanding of the Hertzsprung-Russell diagram (HRD) to put a prior on the available stellar parameter space, and simultaneously infers extinction, effective temperature and distances to stars, based on broadband photometry and Hipparcos parallaxes. \citet{Hanson2014} expand this method to use SDSS and UKIDSS data when parallaxes are absent and also to infer the extinction parameter at high Galactic latitudes.

\citet{Sale2014} use a hierarchical Bayesian system developed in \citet{Sale2012} applied to IPHAS data to map extinction in the northern Galactic plane.

\citet{Green2014} and \citet{Schlafly2014b} combine Galactic priors to obtain probabilistic three dimensional extinction estimates for most of the Galaxy above declination $-30$ degrees with Pan-STARRS1 data. \citet{Vergely2010} and \citet{Lallement2014} apply an inversion method to data from multiple surveys to map the local interstellar medium in particular.

In \citet{Hanson2014} we demonstrated the method used in the present work on SDSS and UKIDSS data of the Galactic poles, finding good agreement with other studies. In this work we use Pan-STARRS1 \citep{Kaiser2010} and Spitzer IRAC data from the GLIMPSE (\textit{Galactic Legacy Infrared Mid-Plane Survey Extraordinaire}) surveys \citep{Churchwell2009,Benjamin2003} to probe the inner few degrees of the Galactic plane, thereby covering more diverse regions of extinction and its variation. This allows us to not only map the line-of-sight extinction but also to quantify the variation of the extinction parameter which characterises the properties and size distribution of dust grains in the interstellar medium.

The paper is organised as follows. In Section~\ref{sec:method} we summarise the method used here, focussing on how we construct the maps presented later on. In Section~\ref{sec:data} we describe the surveys and data products we use to construct the map. Results are presented in Section~\ref{sec:extinctionmap}, where we illustrate the performance and validity of our results. We close with a conclusion and discussion in Section~\ref{sec:conclusion}, suggesting future steps and goals. The map data are available via the \text{MNRAS} website.

\section{Method}
\label{sec:method}
We outlined our Bayesian approach to infer the astrophysical parameters (APs) of individual stars in \citet{Hanson2014}. However, section 2.2 of that article may be a little confusing, as we used the absolute magnitude, $M$, in the description, rather than $\Delta$ (which is the actual parameter), which we use as a proxy for distance modulus $\mu$. This resulted in equation 5 being erroneous (the implementation was correct). We use the same method here, but now provide a more accurate description. It generalises the method introduced by Bailer-Jones (2011) to include distance modulus via a proxy.

We want to obtain the posterior distribution over the parameters given the data and assumptions. The parameters are the monochromatic extinction, $A_0$, the effective temperature, $T$, and the distance modulus, $\mu$. (We will add to this the relative extinction, $R_0$, later.) However, to avoid having to model the dependency of distance modulus on extinction, we instead actually infer $\Delta = m_{\rm r} - M_{\rm r}$. When reporting results we compute distance modulus as $\mu = \Delta - A_{\rm r} = m_{\rm r} - M_{\rm r} - A_{\rm r}$, where $A_{\rm r}$ is calculated as a function of $A_0$ and $T_{\rm eff}$.

The data are the set of colours, the vector $ \bmath{p}$, and the apparent magnitude in one band, $m$.\footnote{We could replace $ \bmath{p}$ and $m$ with the individual magnitudes, but it's methodologically preferable to separate out the spectral and distance information.} $H$ stands for the Hertzsprung--Russell diagram, which introduces our prior knowledge of stellar structure and evolution. This is a two-dimensional probability distribution over $(M, T)$, which we will specify in Section~\ref{sec:data-hrd}. Using Bayes' theorem, the posterior distribution can be written as the product of a likelihood and a prior (multiplied by a normalisation constant $Z^{-1}$) 
\begin{eqnarray}
P(A_0, T, \Delta |  \bmath{p}, m, H) \,=\, \frac{1}{Z} P( \bmath{p}, m | A_0, T, \Delta, H) P(A_0, T, \Delta | H) \ .
\label{eqn:bayes1}
\end{eqnarray}
Using the law of joint probabilities, and the fact that $ \bmath{p}$ is independent of $m$, $\Delta$, and $H$ once conditioned on $A_0$ and $T$, we can write the likelihood as
\begin{eqnarray}
P( \bmath{p}, m | A_0, T, \Delta, H) = P( \bmath{p} | A_0, T) P(m | A_0, T, \Delta, H) \ .
\label{eqn:bayes2}
\end{eqnarray}
The second term can be written as a marginalisation over $M$
\begin{eqnarray}
\begin{aligned}
& P(m | A_0, T, \Delta, H)  \nonumber \\ 
& = \int_M P(m | M, A_0, T, \Delta, H)P(M | A_0, T, \Delta, H) \,{\rm d}M \nonumber \\
& = \int_M P(m | M, \Delta) \frac{P(M, T | H)}{P(T | H)} \, {\rm d}M
\label{eqn:bayes3}
\end{aligned}
\end{eqnarray}
where conditional independence allows us to remove $A_0$, $T$ and $H$ from the first term under the integral.  This is because $m = \Delta + M$, by definition.  Note that because $m$ and $\Delta$ are measured -- and therefore noisy -- quantities, $P(m | M, \Delta)$ is not a delta function. We also removed $A_0$ and $\Delta$ from the second term, because given the HRD and $T$, the distribution over $M$ is fully defined. Note that the right-hand-side no longer has any dependence on $A_0$.  $m$ is conditionally independent of $A_0$ because $H$ and $T$ specify a distribution over $M$, which together with $\Delta$ specifies a distribution over $m$.

Finally, if the prior is separable such that we can write 
\begin{eqnarray}
P(A_0, T, \Delta | H) = P(A_0, \Delta) P(T | H) \ ,
\label{eqn:bayes4}
\end{eqnarray}
then substituting equations \ref{eqn:bayes2}, \ref{eqn:bayes3} and \ref{eqn:bayes4} into \ref{eqn:bayes1} gives
\begin{eqnarray}
\begin{aligned}
& P(A_0, T, \Delta |  \bmath{p}, m, H) = \nonumber \\  
& \frac{1}{Z} P( \bmath{p} | A_0, T) P(A_0, \Delta) \int_M P(m | M, \Delta) P(M, T | H) \, {\rm d}M \ .
\end{aligned}
\end{eqnarray}
This expression is the product of three terms. The first is the probability of measuring the colours given the relevant parameters. The second is the prior over extinction and $\Delta$. The third is an integral over the unknown absolute magnitude, constrained by the HRD and the relationship between $m$, $M$, and $\Delta$. We can generalise the equation to include $R_0$ by simply replacing $A_0$ with $(A_0, R_0)$.

For $A_0$ and $R_0$ we adopt uniform priors over the parameter ranges we explore and zero outside. In practice we only process results further which have estimated APs in the ranges from $\unit{3100-9900}{K}$ and $2.2-5.8$ for $T_{\rm eff}$ and $R_0$, respectively. This is by design, as our HRD prior limits the effective temperature range and $R_0$ is not expected to exceed the extreme values of the above range. Although we do not explicitly limit the range of $A_{\rm 0}$ during the inference, in practice we flag any stars that have an estimated $A_{\rm 0}$ above approximately seven magnitudes, as these stars tend not to fit into our model and we therefore do not trust the inferences. This is typically the case when one or several estimated APs lie at the boundaries of their respective parameter range (which is $\unit{10}{mag}$ for $A_0$). In any case, due to the brightness limits of the surveys and the use of optical photometry, we do not expect to find many stars at very high extinctions and also don't expect to be able to estimate their parameters accurately. For $\Delta$, we also adopt a uniform prior.

\subsection{Extinction}
\label{sec:method-extinction}
In the model we use the monochromatic extinction $A_0$, which only depends on properties of the interstellar medium along the line of sight. Other parameterisations are functions of the star's spectral energy distribution (\textit{SED}) and therefore depend on the effective temperature. We use the extinction curves from \citet{Fitzpatrick1999}, which allow us to vary the extinction parameter $R_0$, which is equivalent to $R_{\rm V}$ in that formulation. We use the same definition of extinction and the extinction parameter as in \citet{Hanson2014}.

\subsection{Forward Model}
\label{sec:method-forwardmodel}
We build a synthetic forward model based on MARCS model spectra \citep{Gustafsson2008} in the temperature range $\unit{2500 - 10\,000}{K}$. Based on the bandpass functions of the survey filters, we compute the absolute photometry for stars with simulated extinction. The zero points are computed in the AB system. We convert the Spitzer IRAC data (which is reported in Vega magnitudes) accordingly.

As the synthetic libraries do not model colours of M dwarfs well, we combine these with empirical stellar loci for the Pan-STARRS1 bands from \citet{Tonry2012} and adapt the synthetic loci at low temperatures ($\unit{\approx3000}{K}$). Synthetic and empirical loci match very well for other spectral types.

For each colour we fit a three-dimensional thin-plate spline to its variation over $A_0$, $R_0$ and $T_{\rm eff}$. These spline models are used to predict the colour for given trial APs, which are compared with the measured colours via the likelihood in Equation~\ref{eqn:bayes1}. To fully model all variations over small parameter changes, we use $\approx 8000$ synthetic stars and allow the splines to have $1000$ degrees of freedom. 
\subsection{Computation}
\label{sec:method-computation}
We use a Metropolis-Hastings Markov Chain Monte Carlo (MCMC) routine to sample the parameter space in logarithmic units of the APs. Using the logarithm forces them to remain positive without the use of an explicit prior to this effect. Sufficient convergence is achieved with $10\,000$ steps each for burn-in and sampling. The sampling steps are of the order of $\unit{0.1}{dex}$ in all variables. To speed up the computation time, we use a lookup table for all parameters. This has a resolution much better than the model accuracy in order to avoid biasing the results from grid effects.

After inferring parameters for all stars, we remove those with parameters at the grid boundaries, resulting in ranges of $\unit{3100 - 9900}{K}$ in $T_{\rm eff}$ and $2.2 - 5.8$ in $R_0$. This post-processing step removes close to $\unit{10}{\%}$ of the stars. In the available dataset these stars have an indicator flag set to $1$ for each affected AP (see Appendix~\ref{app:data}).

\section{Data}
\label{sec:data}
Our extinction map is based on Pan-STARRS1 (PS1) and Spitzer photometry. We crossmatch PS1 and Spitzer IRAC $\unit{3.6}{\mu m}$ point source data from the Glimpse surveys using the API of the cross-match service provided by CDS, Strasbourg\footnote{\url{cdsxmatch.u-strasbg.fr/xmatch}} with a $\unit{1}{''}$ search radius. This results in a data set with $19\,885\,031$ stars. Details on the surveys and data selection are noted below.

\subsection{Pan-STARRS1}
\label{sec:data-ps1}
The Pan-STARRS1 survey has observed the entire sky north of declination $\unit{-30}{^\circ}$ in five filters \citep{Stubbs2010,Tonry2012}. These cover the wavelength range $\unit{400 - 1000}{nm}$. The resulting global photometric calibration is better than $\unit{1}{\%}$ \citep{Schlafly2012}.

We select all point sources classified as stars that have good observations in the five bands, $g_{\rm P1}$, $r_{\rm P1}$, $i_{\rm P1}$, $z_{\rm P1}$ and $y_{\rm P1}$, using the epoch-averaged photometry in each band. We use data collected up to February 2013. We do not take into account any variability observed across multiple epochs. $\unit{90}{\%}$ of stars have $g_{\rm P1}$-band magnitudes between $\unit{16.19 - 21.95}{mag}$. Only a tiny fraction of the stars have photometric uncertainties worse than $\unit{0.1}{mag}$, the median uncertainties in the five bands are in the range $\unit{0.01 - 0.02}{mag}$.

\subsection{Spitzer GLIMPSE}
\label{sec:data-spitzer}
The Spitzer Space Telescope Legacy program GLIMPSE consists of four separate surveys (I, II, 3D, 360), which together cover most of the Galactic plane within a few degrees in latitude. The Infrared Array Camera \citep[IRAC;][]{Fazio2004} is used to image at $3.6$, $4.5$, $5.8$, and $8.0 {\rm \mu m}$. We use only the $3.6 {\rm \mu m}$ data, as the longer wavelength measurements do not improve our parameter estimation. We select point sources that have signal to noise ratios greater than $3$ and closed source flags (\texttt{csf}) of $0$, indicating that no other sources are within $\unit{3}{''}$ of a source. This is to ensure that sources are extracted reliably. The $\unit{90}{\%}$ quantile for $3.6 {\rm \mu m}$-band magnitudes is $\unit{11.20 - 16.47}{mag}$. 

\subsection{Hertzsprung-Russell Diagram}
\label{sec:data-hrd}
\begin{figure}
\includegraphics[width=\columnwidth]{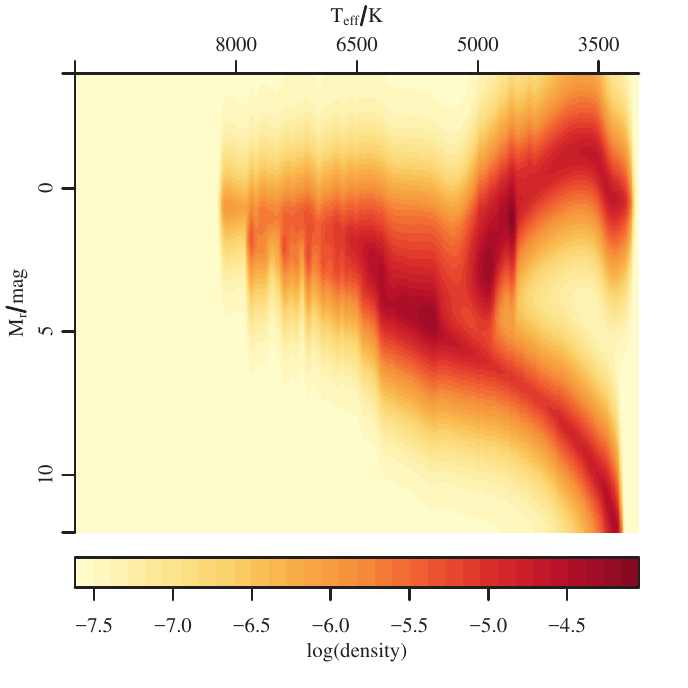}
\caption{Density representation of the HRD, where the integrated probability is normalised to one. The colour scale shows base $10$ logarithm of the density, dark red is high number density and light yellow is low. Light yellow areas denote regions of the parameter space with initially zero probability. A small offset is added to each point before normalisation to avoid this in the actual computation. In this case the offset is approximately $10^{-4}$ times the maximum density, resulting in a value of $-7.5$ in logarithmic density.}
\label{fig:data-hrd}
\end{figure}

As in \citet{Hanson2014}, we use a HRD prior as a constraint in the $T_{\rm eff} - M_{\rm r}$ -plane. To fully account for the distribution in stellar types expected in the Galactic plane, in particular K and M dwarfs, as well as giants, we use the Dartmouth Stellar Evolution Database \citep{Dotter2008}. For fixed solar metallicity, we smooth the data in the HRD plane using a binned kernel density estimate with bandwidths of $\unit{25}{K}$ and $\unit{0.125}{mag}$ in $T_{\rm eff}$ and $M_{\rm r}$, respectively. The temperature range is from $\unit{2500-10\,000}{K}$, the absolute magnitudes vary from $\unit{-4}{mag}$ to $\unit{12}{mag}$. The resulting grid has the pixel dimensions of $751 \times 600$ (as $T_{\rm eff} \times M_{\rm r}$). Before normalisation, a small, non-negative offset is added to all pixels to account for the fact that the regions that are empty in the Dartmouth model HRD in reality may not have exactly zero probability. We show a representation of the HRD in Figure~\ref{fig:data-hrd}.

The HRD of course depends on the metallicity, and as demonstrated in \citet{Hanson2014} the choice affects the results. Unsurprisingly, it is not possible to also estimate metallicity from our photometric data (due in part to the large - a priori unknown - range of $T_{\rm eff}$ and $A_0$ in the data). If we fixed the metallicity of the HRD to a single value, we would obtain artificially precise (but not necessarily more accurate) results for the inferred parameters. To avoid this, we took an HRD and then smoothed it (using a kernel density estimation method). This produces a smooth but conservatively broad HRD; it is broader than the one used in \citet{Hanson2014}. As demonstrated in that paper, the lack of a metallicity determination will be the main limiting factor on the distance accuracy, while the extinction, extinction parameter, and effective temperature are less influenced by this. We make this compromise of a simple HRD as we do not wish to introduce yet more dependencies by imposing a complex Galaxy model.
%

\section{Maps}
\label{sec:extinctionmap}
We apply our method to the cross-matched PS1-GLIMPSE photometry to obtain individual AP estimates for all stars individually. To summarise and visualise these results we bin stars with a fixed angular resolution of $\unit{7}{'} \times \unit{7}{'}$ in $l$ and $b$. We present the maps after converting the extinction values to the $r$-band extinction $A_{\rm r_{\rm PS1}}$ \citep[see][for details of this conversion]{Hanson2014}. To compute the variation in extinction $A_{\rm r}$ (and $R_0$) along the line-of-sight at any value of distance modulus $\mu_j$ we calculate the weighted mean extinction $\left< A_{\rm r} \right>_j$ and standard deviation $\Sigma_j$ (and analogously for $R_0$) for all stars in a single bin which have a distance modulus estimate within one magnitude of our selected position. These are
\begin{eqnarray}
\left< A_{\rm r} \right>_j &=& \frac{\sum A_{{\rm r}, i} w_{i, j}}{\sum w_{i, j}} \quad , \nonumber \\
\Sigma_j &=& \sqrt{\frac{\sum w_{i, j} (A_{{\rm r}, i}-\left< A_{\rm r} \right>_j)^2}{\frac{N-1}{N}\sum w_{i, j}}} \quad ,
\label{eq:extinctionmap-weightedmean}
\end{eqnarray}

where the sums are over $i$. The weight $w_{i, j}$ is a measure of the difference between the inferred stellar distance modulus $\mu_i$ and the cell distance $\mu_j$. The confidence intervals about the mode are not symmetric, so we use a split Gaussian to approximate the distribution they describe. For each star we compute the weight using the asymmetric Gaussian (or split normal distribution), parameterised by the mode and the standard deviations, $\sigma_1$ and $\sigma_2$, of each half of the Gaussian:
\begin{equation}
w_{i, j} = \frac{2}{\sqrt{2\pi}(\sigma_1+\sigma_2)}\cdot\exp\left( -\frac{(\mu_i - \mu_j)^2}{2\sigma_k^2}\right) \ ,
\label{eq:extinctionmap-weights}
\end{equation}
In the case when $\mu_i$ is smaller than $\mu_j$ then $\sigma_k= \sigma_1$, otherwise  $\sigma_k = \sigma_2$. This is a convenient and fast substitute for summing over all the 2D PDFs we obtain from the inference. Stars with small confidence intervals are weighted more strongly than those with large ones. This procedure can be applied to any arbitrary distance modulus step $\mu_j$. This is repeated for every angular bin to construct a full three dimensional representation of the cumulative line-of-sight extinction. Analogously we use the same procedure with the extinction parameter $R_0$, allowing us not only to follow the extinction variation along the line of sight, but also to look at the properties of the dust. Due to the selection process, it is in principle possible that individual stars appear in two consecutive distance bins, indicating that this measure is similar to a running (weighted) mean. For each cell we require at least $10$ stars to compute the result.

We use distance modulus as the distance variable because it straightforwardly captures the uncertainty which increases with distance. For example, the relative error in distance for a distance modulus error of $\delta \mu = \unit{1}{mag}$ at $d = \unit{1}{kpc}$ is $\delta d = \unit{0.46}{kpc}$, whereas at $d = \unit{5}{kpc}$ it increases to $\delta d = \unit{2.3}{kpc}$. It is important to note that although the uncertainty in $\mu$ may be symmetric, it will not be in $d$.

\begin{figure}
\includegraphics[width=\columnwidth]{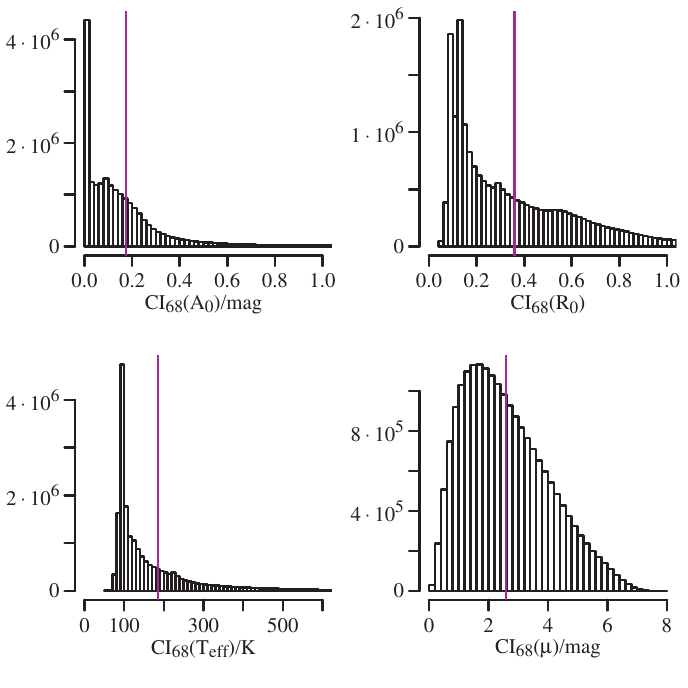}
\caption{Histograms of the widths of the $68\%$ confidence intervals of extinction $A_0$, extinction parameter $R_0$, effective temperature $T_{\rm eff}$ and distance modulus $\mu$. The purple vertical lines indicate the mean values in each case. For $A_0$ this corresponds to $\unit{0.17}{mag}$, for $R_0$ it is $\unit{0.36}{}$, for $T_{\rm eff}$ it is $\unit{185}{K}$ and for $\mu$ it is $\unit{2.6}{mag}$.}
\label{fig:data-errhist}
\end{figure}

The mean uncertainties on extinction $A_{\rm 0}$, extinction parameter $R_0$, effective temperature $T_{\rm eff}$ and distance modulus $\mu$, based on the widths of the $68\%$ confidence intervals of the individual stellar parameter estimates, are $\unit{0.17}{mag}$, $0.36$, $\unit{185}{K}$ and $\unit{2.6}{mag}$, respectively. For each star we obtain an entire PDF over the parameters, from which we compute the confidence intervals. The lower bound of the $68\%$ confidence interval has $16\%$ of the probability below it, whereas the upper bound has $16\%$ of the probability above it. Histograms of the uncertainty distributions are shown in Figure~\ref{fig:data-errhist}.

\begin{figure}
\includegraphics[width=\columnwidth]{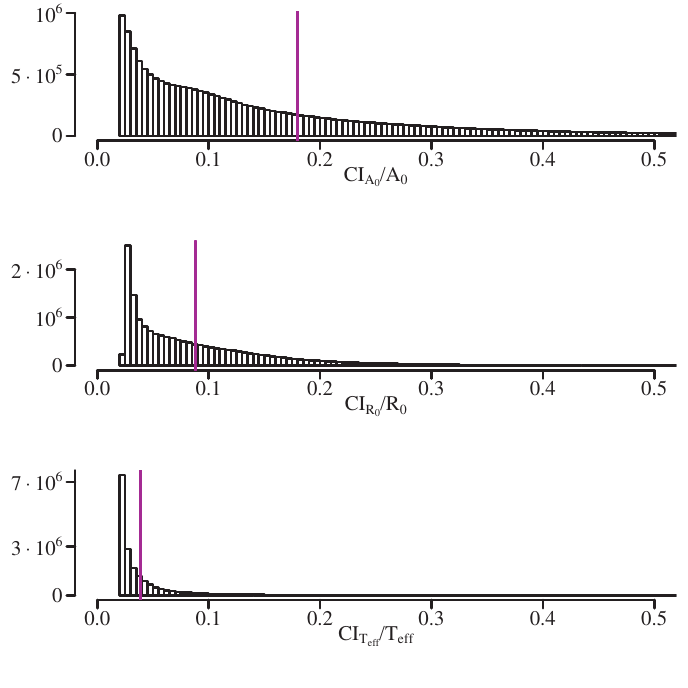}
\caption{Histograms of relative uncertainties as defined by the widths of $68\%$ confidence intervals divided by mean for $A_0$, $R_0$ and $T_{\rm eff}$. The purple vertical lines indicate the mean values in each case. For $A_{\rm 0}$ this corresponds to $0.17$, for $R_0$ it is $0.09$ and for $T_{\rm eff}$ it is $0.04$.}
\label{fig:relerror}
\end{figure}

In Figure~\ref{fig:relerror} we show histograms of the relative uncertainties for the APs for each star (distance modulus is not included, as it is a fractional distance.) These are computed by dividing the width of the $68$ per cent confidence intervals by the mean. The mean relative uncertainties are $0.17$, $0.09$ and $0.04$ for extinction, extinction parameter and effective temperature, respectively.

\begin{figure*}
\includegraphics[width=\textwidth]{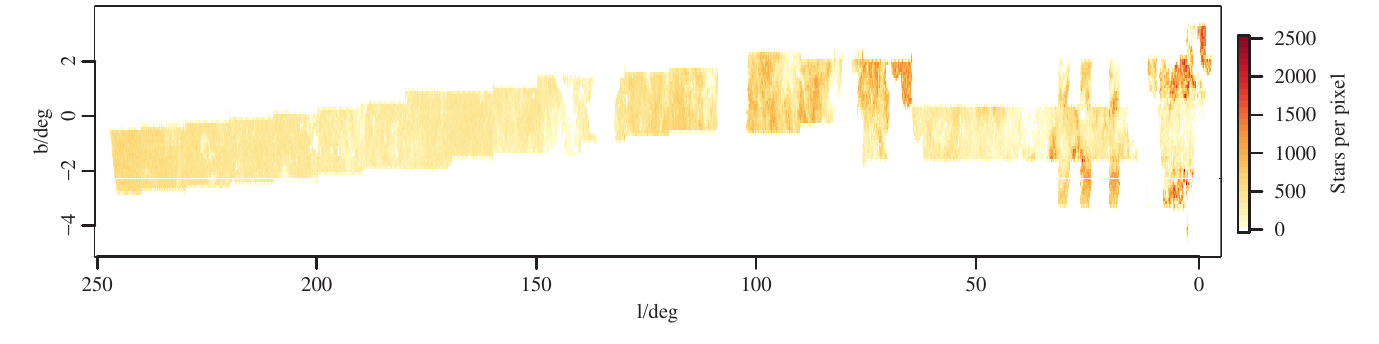}
\caption{Stars per $\unit{7}{'} \times \unit{7}{'}$ pixel across the data footprint shown as a colour density scale (white areas are not covered by the data). The mean density is $400$ stars per pixel. The scale is limited to $2500$ stars per pixel (red), and any pixels with more stars are shown in this colour.}
\label{fig:extinctionmap-counts}
\end{figure*}

In Figure~\ref{fig:extinctionmap-counts} we illustrate the density of stars per pixel for each line of sight. Note that this does not indicate directly how many stars are used at each distance slice. We impose minimal requirements in this case (see above). The mean density is nearly $400$ stars per $\unit{7}{'} \times \unit{7}{'}$ pixel, whereby some pixels have only a few stars (not counting regions not covered by the data set). The maximum is $2\,931$, the most dense pixels tend to be situated slightly above and below $b = 0$ around the Galactic centre. As expected the density decreases as we move away from the Galactic centre in longitude.

\subsection{Extinction $A_0$}
\label{sec:extinctionmap-ar}
\begin{figure*}
\includegraphics[width=\textwidth]{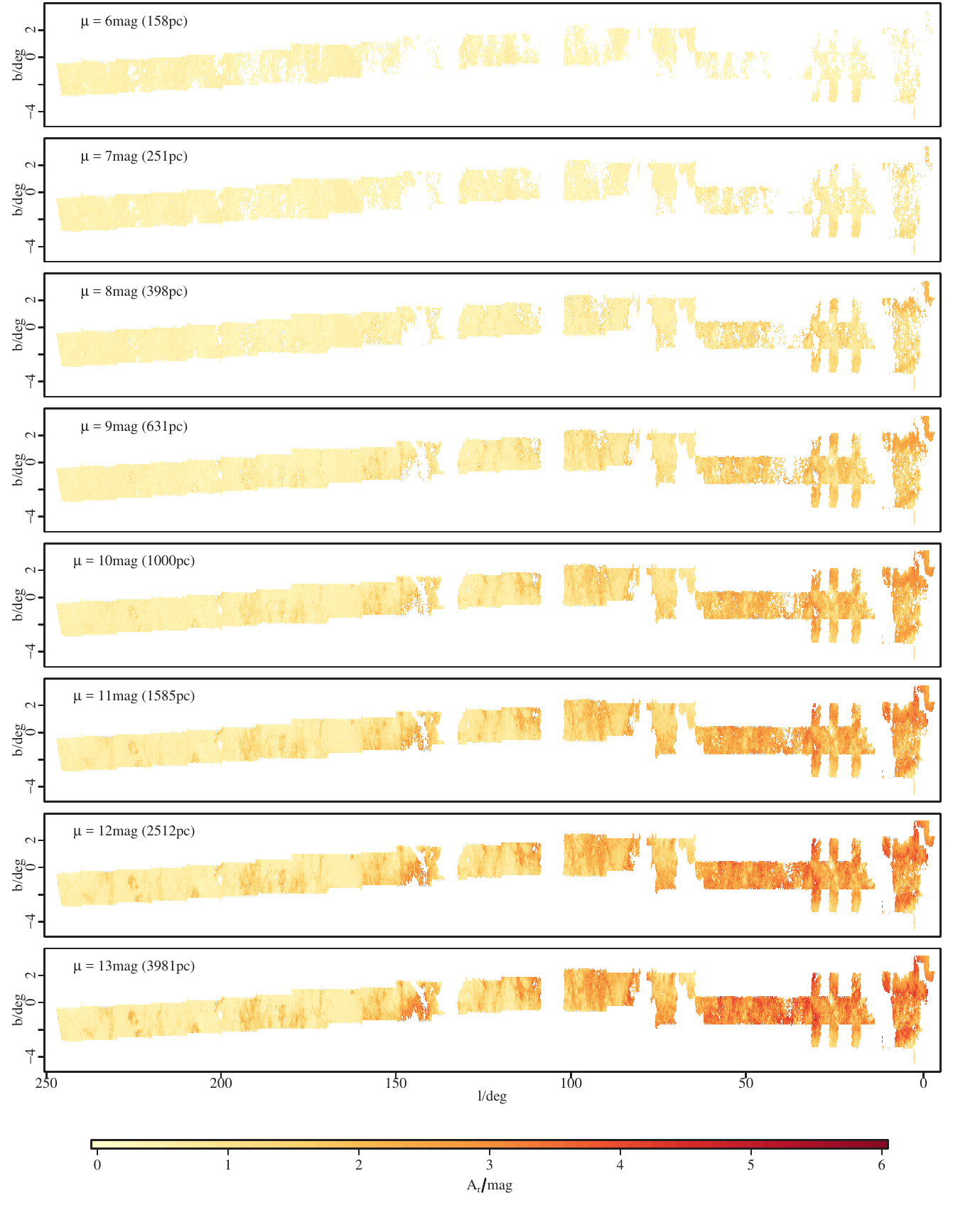}
\caption{Cumulative line of sight extinction at distance moduli from $\mu = \unit{6-13}{mag}$ in $r_{\rm P1}$ -band. White regions are either not covered by the data footprint or (particularly at closer distances) do not contain a sufficient number of stars.}
\label{fig:extinctionmap-ar}
\end{figure*}

Figure~\ref{fig:extinctionmap-ar} shows the cumulative line of sight extinction for eight distance slices from $\mu = \unit{6-13}{mag}$ in units of $r_{\rm P1}$ -band extinction as two-dimensional slices of the full map through the Galactic plane. Various structures are visible. In particular the lack of higher extinctions between $l = 100-150^{\circ}$ and towards larger distances coincides with the warp in the dust distribution noted by \citet{Marshall2006} and \citet{Sale2014}. In Section~\ref{sec:extinctionmap-validation} we will analyse in more detail a few particular molecular clouds, which we will also use to validate the overall method.

At closer distances some cells contain insufficient stars to be assigned an extinction estimate and therefore appear white. The colour scale is limited to $A_{\rm r}$ $\leq\unit{6}{mag}$; the highest extinction estimate for any pixel is $A_{\rm r}$$=\unit{5.2}{mag}$, although individual stellar estimates may be larger.

\begin{figure}
\includegraphics[width=\columnwidth]{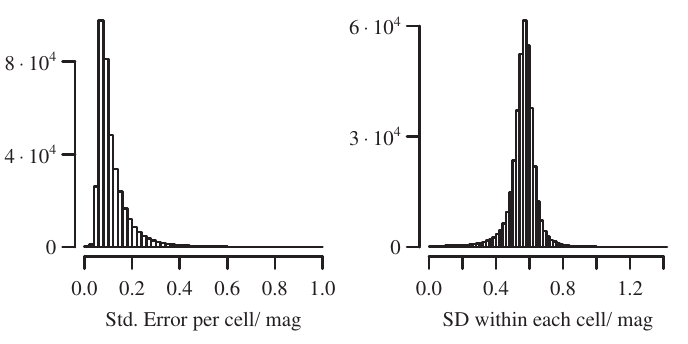}
\caption{Left panel: Histogram of the predicted uncertainty of distance modulus. The standard error is computed for each cell, using the differences of all inferred distance moduli and the fixed cell distance. The mean of this distribution is $\unit{0.12}{mag}$. Right panel: Distribution over all cells of the standard deviation of distance moduli within each cell. This distribution has a mean of $\unit{0.56}{mag}$. In both panels, cells with distance moduli $\mu = \unit{6 - 13}{mag}$ are included.}
\label{fig:stderror}
\end{figure}

Based on the distribution of the standard deviation of individual stellar distances within the three dimensional cells and the standard error of the mean in each cell (per angle and distance, for which a summary is shown in Figure~\ref{fig:stderror}), we estimate that distances are only reliable from $\mu = \unit{6 - 13}{mag}$. At closer distances we observe few to no stars due to the bright magnitude limits of the surveys. Beyond the upper limit, distance uncertainties become very large and the distance estimates themselves are no longer useful (see the relation between distance modulus and distance uncertainties above). Those distance slices are not presented here (although the individual stellar distances are available in our published data set).

The predicted uncertainty is illustrated by the distribution of the model-predicted standard errors in the distance modulus and is shown in the left panel of Figure~\ref{fig:stderror}. For each cell, we compute the standard error of all inferred distance moduli from the fixed cell distance. The average of these is $\unit{0.12}{mag}$ with a standard deviation of $\unit{0.08}{mag}$. The distribution over all cells of the standard deviation of distance moduli within each cell is shown in the right panel of Figure~\ref{fig:stderror}. The distribution has a mean of $\unit{0.56}{mag}$ and standard deviation of $\unit{0.09}{mag}$. The results indicate that the selected distance slices represent the underlying distance distribution of the stars well.

\begin{figure*}
\includegraphics[width=\textwidth]{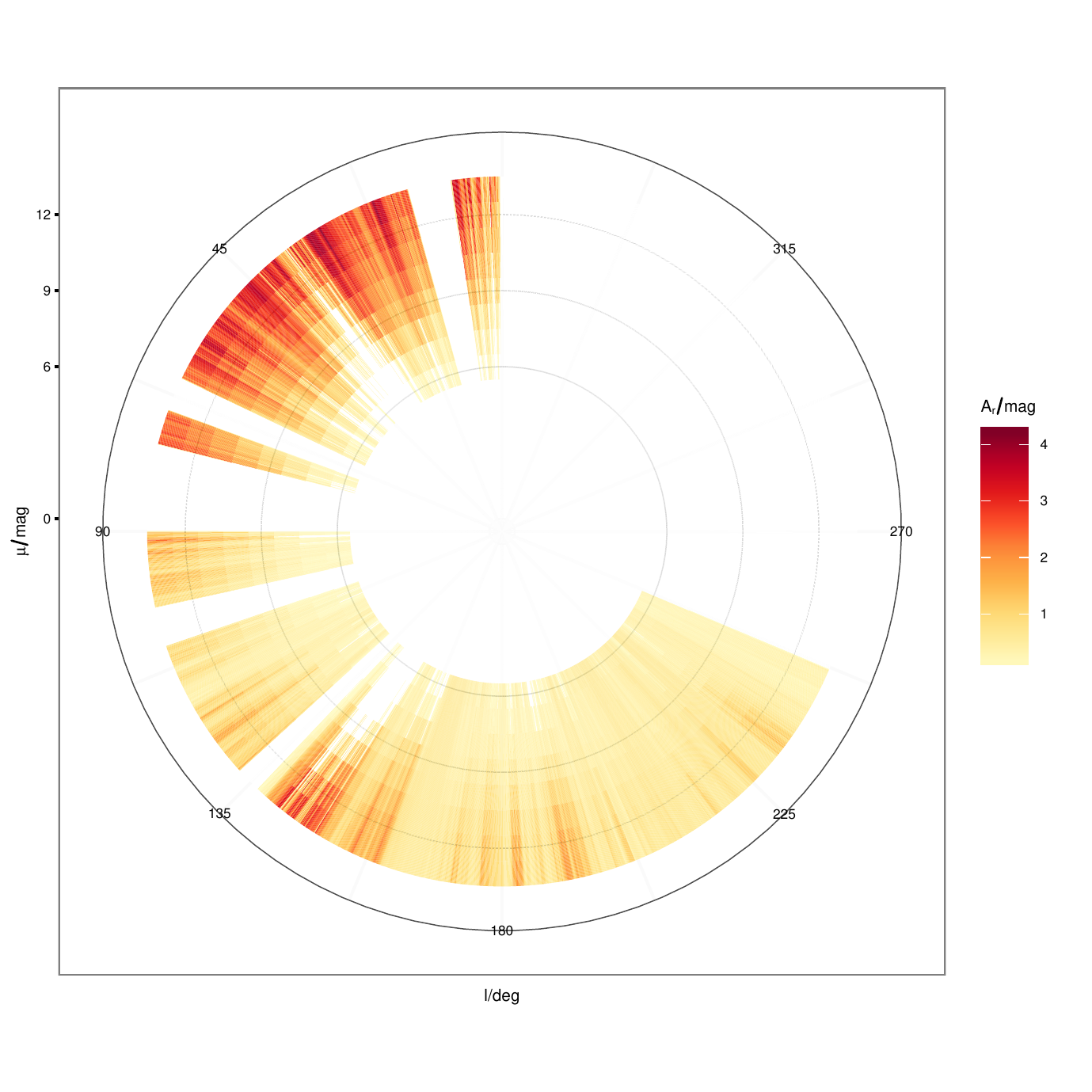}
\caption{Projected extinction map with vertical extent of the Galactic plane from $b = \unit{-0.21}{^\circ} - \unit{+0.25}{^{\circ}}$ in which the five central latitude slices are averaged. The Sun is at the centre of the diagram. The distance moduli on the left edge refer to the radii of the circles. The physical distances of the indicated distance moduli $6$, $9$ and $\unit{12}{mag}$ are $158$, $631$ and $\unit{2512}{pc}$, respectively.}
\label{fig:extinctionmap-topdown}
\end{figure*}

In Figure~\ref{fig:extinctionmap-topdown} we show a top-down view of the Galaxy at $b = 0$ in which we average over the five central latitude slices, i.e. from $b = \unit{-0.21}{^\circ}$ to $b = \unit{0.25}{^{\circ}}$. As a reference, a distance modulus of $\unit{5}{mag}$ ($\unit{10}{mag}$) is equivalent to a distance of $\unit{100}{pc}$ ($\unit{1000}{pc}$). Here we can clearly see the $\unit{1}{mag}$ length of the distance modulus slices as well as the expected increase of extinction within a few ${\rm kpc}$ towards the Galactic centre at the top of the figure. As the measured extinction in neighbouring cells are only correlated in the radial direction, but not in longitude (or latitude), many discontinuities can be seen. 

\subsection{Extinction Parameter $R_0$}
\label{sec:extinctionmap-r}
\begin{figure*}
\includegraphics[width=\textwidth]{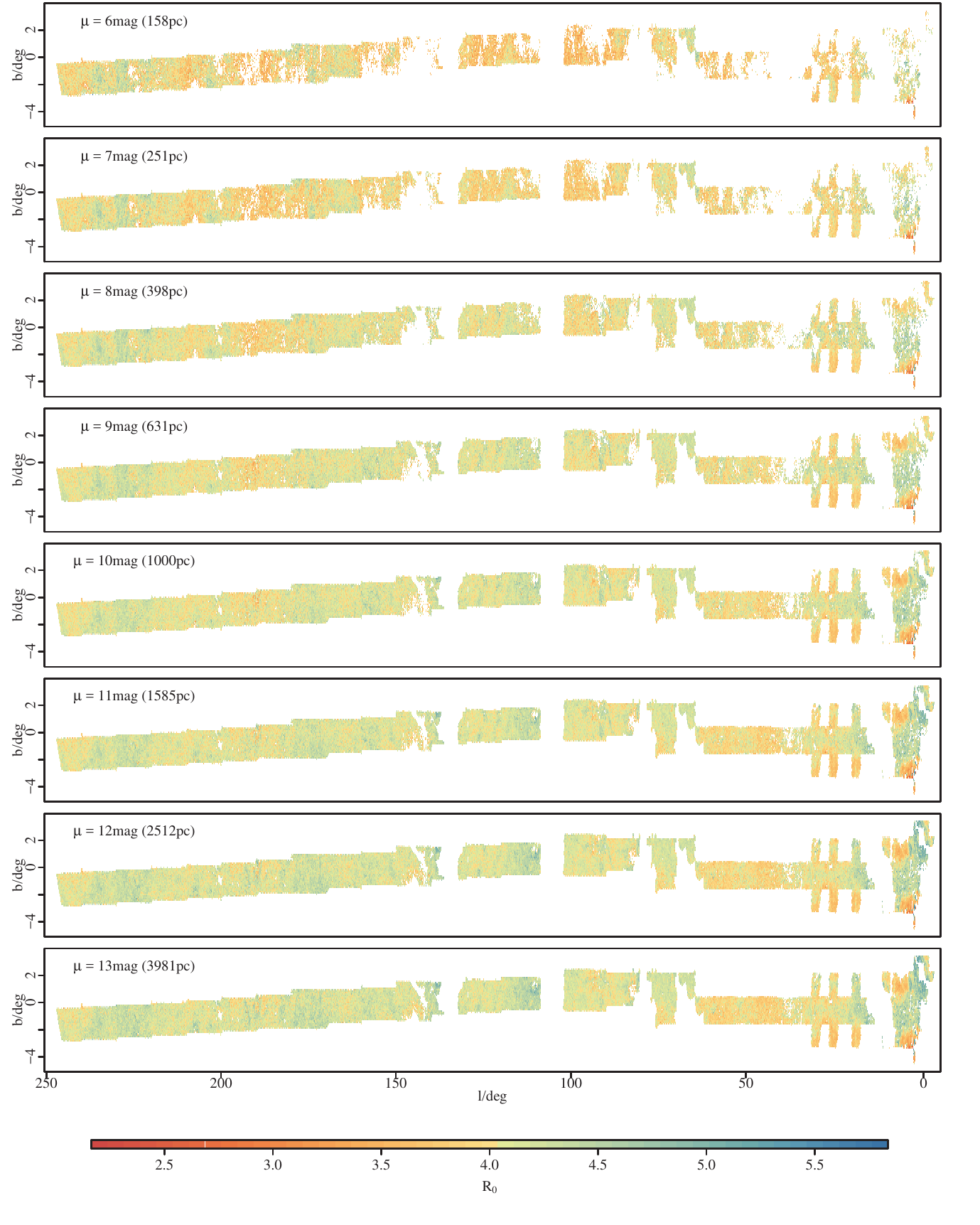}
\caption{Extinction parameter $R_0$ at distance moduli from $\mu = \unit{6-13}{mag}$, computed according to Equation~\ref{eq:extinctionmap-weightedmean} (weighted mean). Again, white regions are either not covered by the data footprint or do not contain enough stars to be assigned a parameter estimate.}
\label{fig:extinctionmap-r}
\end{figure*}

As mentioned in Section~\ref{sec:method} we not only infer extinction $A_{\rm 0}$ but also the extinction parameter $R_0$. In Figure~\ref{fig:extinctionmap-r} we show this parameter in slices of distance modulus, analogously to Figure~\ref{fig:extinctionmap-ar}. It is clear that variations here follow those in extinction. Although there is an indication that in some regions with higher extinction $R_0$ increases above the mean of $\left< R_{\rm 0} \right > = 4.1 \pm 0.27$, we do not detect a global correlation between the two parameters. Only for the two closest distance slices and for low extinctions ($A_{\rm r} < \unit{0.5}{mag}$), is there an inkling that $R_0$ increases with $A_0$. Whilst we trust the variations of $R_0$ we measure, we are less certain about the absolute values. This again has to do with model uncertainties and parameter degeneracies that we are unable to remove. Both extinction and the extinction parameter are cumulative along the line of sight to any given distance. All the dust along the line of sight contributes to any individual estimate. Because of this, correlations between these two cumulative parameters are harder to see: at larger distances, the length scale over which the dust properties are averaged increases. For both the $A_0$ and $R_0$ estimates we use only stars in a limited distance range around the specified distance.

Our results show that the extinction law is not universal. This has previously been asserted by other authors, such as \citet{Gao2009} and \citet{Chen2013} who also look at the variation in large regions of the Galaxy.

\begin{figure}
\includegraphics[width=\columnwidth]{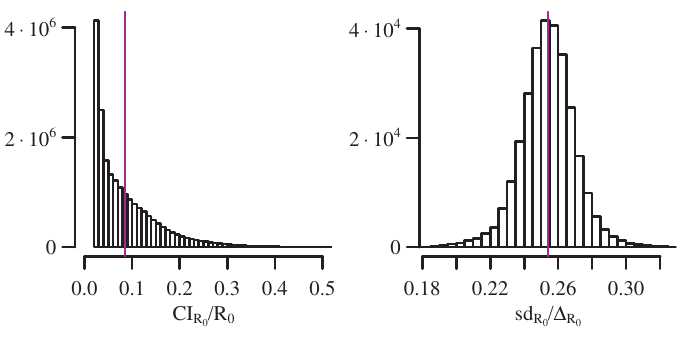}
\caption{Left panel: Relative uncertainty of the $R_0$ estimates for all individual stars. The value denotes 68 per cent confidence interval over the mean inferred parameter. The purple line indicates the mean of the distribution at $0.085$. Right panel: Histogram of the standard deviation relative to the range of $R_0$ estimates of all cells. The mean of $0.25$ is indicated by the purple line.}
\label{fig:extinctionmap-validation-rhist}
\end{figure}

The estimates of $R_0$ for individual stars have, on average, an uncertainty of about $\unit{10}{\%}$, as characterised by the ratio of the width of the confidence interval to parameter estimate. This is shown in the left panel of Figure~\ref{fig:extinctionmap-validation-rhist} as a histogram of all stars. The right panel illustrates the accuracy of the average $R_0$ estimates from Figure~\ref{fig:extinctionmap-r}. We compute the ratio of the standard deviation to the range of $R_0$ for the stars contained in each cell. This average is $0.25$ and indicates that for any individual cell the mean $R_0$ estimate is well constrained, despite possible variations arising from the fact that APs are inferred for all stars individually.

\begin{figure}
\includegraphics[width=\columnwidth]{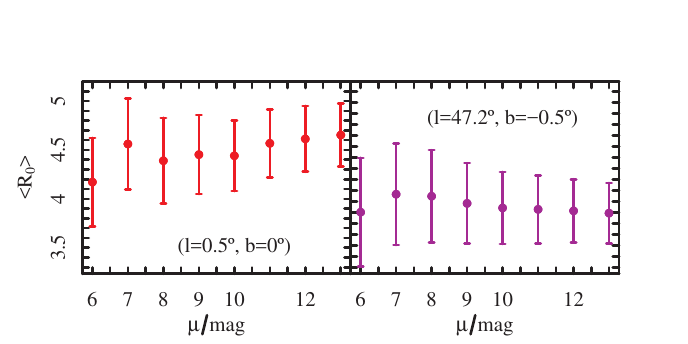}
\caption{Distance modulus versus mean extinction parameter of cells within half a degree, centred at $l=0.5^{\circ}, b=0$ (left panel) and $l=47.2^{\circ}, b=-0.5^{\circ}$.}
\label{fig:extinctionmap-validation-prof}
\end{figure}

\citet{Zasowski2009} find that the inner fields of the Galaxy correspond to a larger $R_0$, whereas outer fields tend to have a lower value. We also find this, as exemplarily shown in Figure~\ref{fig:extinctionmap-validation-prof} where we plot the average extinction parameter, $\left<R_0\right>$ over several cells as a function of distance modulus for two different lines of sight. The first (left panel) is centred on $l=0.5^{\circ}, b=0$ towards the Galactic centre. The second (right panel) is centred on $l=47.2^{\circ}, b=-0.5^{\circ}$. In both cases we average over approximately half a degree in $l$ and $b$, corresponding to 5 pixels in each direction at our resolution. We immediately see that the inner profile increases towards the Galactic centre, above the average of $4.1$ for our data, an effect that is also seen by \citet{Gontcharov2012}.

The profile for the outer field, which we expect to look through more diffuse dust, remains basically flat at a value below the global average. The mean extinction parameter for this line of sight has a value of $3.9 \pm 0.37$, very close to the value of $3.8 \pm 0.20$ we find in section $5.3$ of \citet{Hanson2014} for regions around the Galactic poles.

Our results for $R_0$ suggest a higher value for the diffuse interstellar medium than previous studies indicate. \citet{Moertsell2013} uses quasar data towards the Galactic poles to find $R_{\rm V} \approx 3$ with a relative uncertainty of $\unit{10}{\%}$. \citet{Savage1979} obtain a value of $3.1$ with a similar uncertainty. However, \citet{Jones2011} find a median value of $3.38$ with at median uncertainty of $0.42$ after fitting SDSS spectra of M dwarfs within $\unit{1}{kpc}$ of the Sun. Their resulting distribution is incompatible with a Gaussian with a width of $\sigma = 0.42$ centered at $3.1$.

We have no reason not to believe our results: we find no systematic errors in the data that could, for example, arise from unexpected correlations between $R_0$ and $A_0$ and/or $T_{\rm eff}$ and thus affect the parameter inference. This is clear from \citet{Hanson2014} where the extinction results for Galactic pole regions are not strongly affected by the inclusion of $R_0$ as an inferred parameter.

\subsection{Validation}
\label{sec:extinctionmap-validation}
To validate our results, in particular the relatively uncertain distances, we compare some of our lines-of-sight with distance estimates to molecular clouds in \citet{Schlafly2014a}, who use Pan-STARRS1 photometry to measure and model distances to high statistical accuracy. From Table 1 in that work we select the clouds whose coordinates lie within our survey limits. These are \textit{CMa OB1} with three individual measurements at $(l,b) = (224.5^{\circ}, -0.2^{\circ})$, $(222.9^{\circ}, -1.9^{\circ})$ and $(225.0^{\circ}, -0.2^{\circ})$, as well as \textit{Maddalena} at $(l,b) = (217.1^{\circ}, 0.4^{\circ})$. The reported distances to these clouds are $1369^{+64}_{-56}$, $1561^{+79}_{-77}$, $1398^{+63}_{-59}$ and $2280^{+71}_{-66}$ ${\rm pc}$, respectively, which in distance modulus are $10.68^{+0.10}_{-0.09}$, $10.97^{+0.11}_{-0.11}$, $10.73^{+0.10}_{-0.09}$ and $11.79^{+0.07}_{-0.06}$ ${\rm mag}$.

\begin{figure}
\includegraphics[width=\columnwidth]{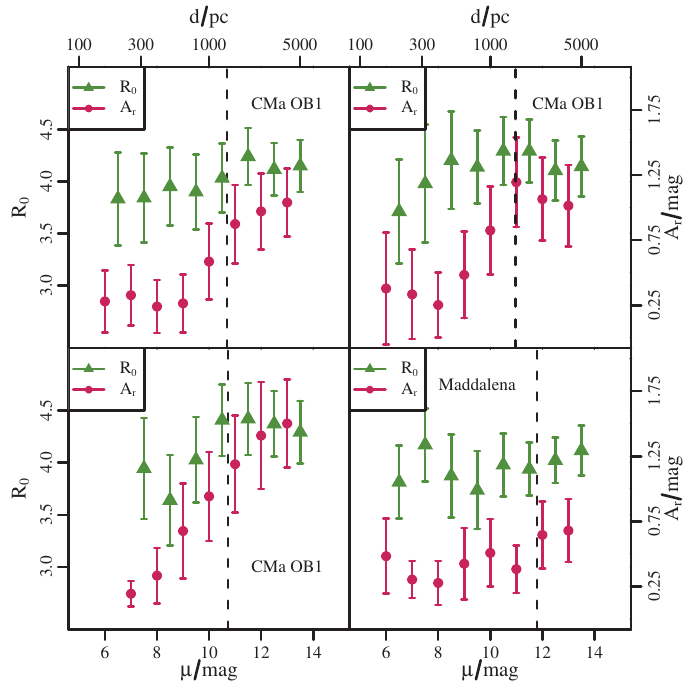}
\caption{Cumulative extinction $A_{\rm r}$ (magenta circles) and extinction parameter $R_0$ (green triangles) as function of distance modulus $\mu$ towards four molecular clouds. See text for the coordinates of the fields. The error bars are computed using Equation~\ref{eq:extinctionmap-weightedmean}. The dashed vertical lines indicate the distances reported in \citet{Schlafly2014a}.}
\label{fig:extinctionmap-validation-profiles}
\end{figure}

In Figure~\ref{fig:extinctionmap-validation-profiles} we show the extinction $A_{\rm r}$ (magenta circles) and extinction parameter $R_0$ (green triangles) as a function of distance modulus for our data using stars within $\unit{7}{'}$ of the coordinates given above. The dashed lines indicate the \citet{Schlafly2014a} distances of the clouds. The mean and error bars are computed according to Equation~\ref{eq:extinctionmap-weightedmean}.

\begin{figure}
\includegraphics[width=\columnwidth]{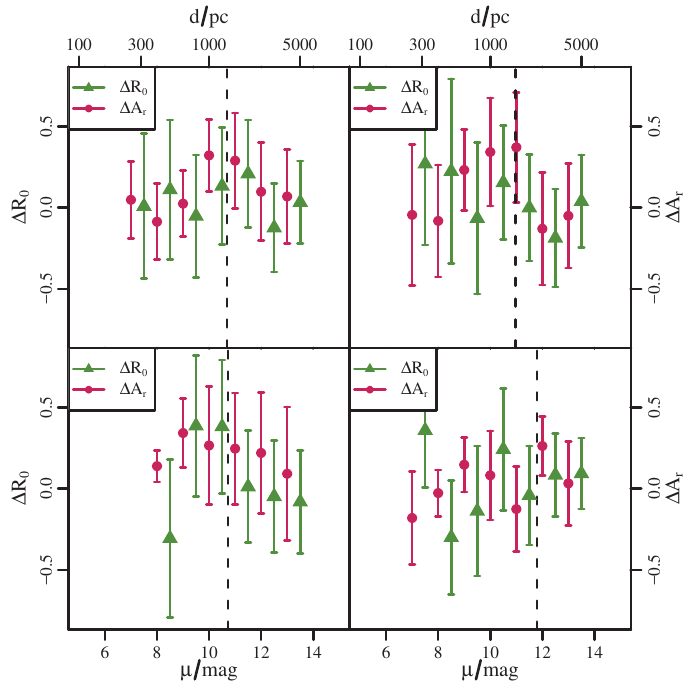}
\caption{Differential extinction $A_{\rm r}$ (magenta circles) and extinction parameter $R_0$ (green triangles) as function of distance modulus $\mu$ towards four molecular clouds. The differentials are computed between distance modulus steps of $\unit{1}{mag}$. See text for the coordinates of the fields. The dashed vertical lines indicate the distances reported in \citet{Schlafly2014a}.}
\label{fig:extinctionmap-validation-profiles-diff}
\end{figure}

Similarly, in Figure~\ref{fig:extinctionmap-validation-profiles-diff} we show differential profiles of $A_{\rm r}$ (magenta circles) and $R_0$ (green triangles), where the values quantify the change in both parameters in steps of $\Delta\mu = \unit{1}{mag}$.

Despite not explicitly measuring distances to individual objects, it is clear that our method manages to capture real features in the extinction distribution. We see that the total extinction $A_{\rm r}$ generally increases around the inferred positions of the clouds, indicating an increase of the underlying dust density around that position. This feature is more pronounced in the two top panels, although the clouds could be responsible for the more gradual increase in extinction in the other two panels as well. This is highlighted in Figure~\ref{fig:extinctionmap-validation-profiles-diff}, where the increase in extinction can be seen more clearly in the top two panels. The interpretation of the bottom two panels in both figures is less clear cut, decspite there being marginal changes in $A_{\rm r}$ and $R_0$ around the literature distances of the clouds. However, the spread in $A_{\rm r}$ (and $R_0$) is generally quite large, and the distances are so uncertain that we are not necessarily confident of having detected the clouds. In all four panels the mean extinction decreases slightly again beyond $\mu = \unit{13}{mag}$. We do not trust values beyond this distance (see Section~\ref{sec:extinctionmap-ar} and Figure~\ref{fig:stderror} for details), as we do not expect to detect many stars at large distances due to the faint magnitude limits of the input catalogues and the resulting selection effects.

The value of the extinction parameter $R_0$ also appears to increase in sync with the increase of extinction, although the magnitude of variation tends to be within the range of uncertainty. Nevertheless, the overall picture is one where there are dense dust clouds which cause the cumulative line of sight extinction to increase above some foreground value. This suggests that the inferred parameters we obtain with our method are trustworthy and physically plausible, at least on a relative scale.

To further probe this, we compare our results with those of \citet{Berry2012} (B12) who combine SDSS and 2MASS data to calculate $A_{\rm r}$ and $R_{\rm V}$ using a straight-forward fit to stellar templates. We take a subset of the common survey area from $l = 49-51^{\circ}$ and $b=-1^{\circ}$ to $b=1^{\circ}$ and compute a 3D dust map based on their results using Equation~\ref{eq:extinctionmap-weightedmean}. Due to the different sensitivities and depths of the surveys we use the distance slices at $\mu =\unit{9}{mag}$ and $\mu =\unit{10}{mag}$ for further comparison, as other distances have many empty cells in one or both data sets. Qualitatively we find similar behaviour and features in the $A_{\rm r}$ extinction map, although the average extinction in our data is $\left< A_{\rm r}\right>=\unit{1.63\pm0.44}{mag}$, whereas the B12 data suggests an average of $\left< A_{\rm r}\right>_{\rm B12}=\unit{2.41\pm0.46}{mag}$. The standard deviations are similar in both cases. For the extinction parameter $R_0$ we obtain an average value of $\left< R_0\right> = 4.04$ with a standard deviation of $0.20$, whereas $\left< R_0\right>_ {\rm B12}= 3.02$ (standard deviation is also $0.20$.) These differences are also reflected when individually cross-matching the stars in the common footprint. The differences (this work minus B12) on average are $\unit{-0.18}{mag}$ for $r$-band extinction and $0.79$ for the extinction parameter. As expected, due to B12's work strongly favouring a value of $3.1$ for a large fraction of stars, we measure a standard deviation of $1.26$ in $R_0$ between the two datasets and see that the differences increase as our $R_0$ estimates increase. Our extinction results agree reasonably well with the previous work. However, we seem to have systematically higher values of $R_0$, which, as discussed previously, may result from fixing the metallicity in the HRD and/or using synthetic spectral templates. Nevertheless, we are much more confident in our relative values of the extinction parameter (and $A_0$), as our model assumptions have much less effect on our ability to measure these.

To exclude the possibility that requiring NIR data could be a cause for the aforementioned differences, we select a random sample of $10\,000$ stars in the same region based purely on their presence in the PS1 data-set. We require no counterpart in the GLIMPSE surveys. Comparing the average widths of the $68\%$ confidence intervals and the average relative uncertainties with results that include GLIMPSE data, as shown in Table \ref{tab:ps1comp}, we find that including the $\unit{3.6}{\mu m}$ photometry significantly improves the precision of the inferred APs. Especially the $R_0$ estimates benefit from the additional band, reducing the average width of the confidence intervals from $0.30$ to $0.22$.

\begin{table}
	\caption{Average widths of $68\%$ confidence intervals $\langle$CI$\rangle$ and mean relative uncertainties $\langle$CI/AP$\rangle$ for extinction $A_0$, effective temperature $T_{\rm eff}$, extinction parameter $R_0$ and distance modulus $\mu$ in the cases of including $\unit{3.6}{\mu m}$ photometry (left) and using only PS1 bands (right).}
 	\label{tab:ps1comp}
 	\begin{tabular}{@{}ccccc@{}}
		\hline
		& \multicolumn{2}{c}{PS1 + $\unit{3.6}{\mu m}$} & \multicolumn{2}{c}{PS1 only} \\
		AP & $\langle$CI/AP$\rangle$ &$\langle$CI$\rangle$ & $\langle$CI/AP$\rangle$ & $\langle$CI$\rangle$ \\
		\hline
		$A_0$ & 0.12 & $\unit{0.23}{mag}$ & 0.14 & $\unit{0.24}{mag}$ \\
		$T_{\rm eff}$ & 0.04 & $\unit{217}{K}$ & 0.04 & $\unit{206}{K}$ \\
		$R_0$ & 0.05 & 0.22 & 0.07 & 0.30\\
		$\mu$ & - & $\unit{2.1}{mag}$ & - &$\unit{2.6}{mag}$ \\
    		\hline
	\end{tabular}
\end{table}

As this sample generally lacks GLIMPSE counterparts, we cannot measure differences in the AP estimates for individual stars. To nevertheless ensure that we have not introduced (or removed) any systematic effects on the inference, we compare the inferred APs for stars from the initial cross-matched sample when including and excluding $\unit{3.6}{\mu m}$ photometry. In this situation the mean differences (including minus excluding $\unit{3.6}{\mu m}$ data) of the APs for these stars are only $\unit{0.07}{mag}$, $\unit{-45}{K}$, $0.01$ and $\unit{0.33}{mag}$ for $A_0$, $T_{\rm eff}$, $R_0$ and $\mu$, respectively. This indicates that the inclusion of the NIR band does not introduce systematic differences, but actually improves the inference.

\section{Conclusion}
\label{sec:conclusion}
We have presented three dimensional maps in cumulative line of sight extinction $A_{\rm 0}$ and extinction parameter $R_0$ which are constructed using a Bayesian method. This method is general and not bound to specific photometric systems. It is based on work by \citet{Bailer-Jones2011} and expanded in \citet{Hanson2014}. We take advantage of the physical understanding of stellar evolution that is encapsulated in the Hertzsprung-Russell Diagram. Using photometric measurements of $19\,885\,031$ stars with data from the cross-matched Pan-STARRS1 and Spitzer Glimpse surveys (six bands in total), we infer extinction $A_{\rm 0}$, extinction parameter $R_0$, effective temperature $T_{\rm eff}$ and distance modulus $\mu$ to all stars individually. We achieve mean relative uncertainties of $0.17$, $0.09$, $0.04$ and $0.18$ for extinction, extinction parameter, effective temperature and distance modulus, respectively whilst obtaining average uncertainties of $\unit{0.17}{mag}$, $0.36$, $\unit{185}{K}$ and $\unit{2.6}{mag}$ for the four parameters. We emphasise that while we believe the $R_0$ variations we measure, we are less confident in the absolute values.

Using these inferred parameters we compute the estimated total extinction to arbitrary distances and estimates of the extinction parameter, as formulated in Equation~\ref{eq:extinctionmap-weightedmean}. The angular stellar density allows us achieve a reliable resolution of $\unit{7}{'} \times \unit{7}{'}$ in latitude and longitude. We select steps of $\unit{1}{mag}$ in distance modulus. From the distribution of distance estimates within all three-dimensional cells, we estimate that the reported extinction map is reliable from $\mu = \unit{6-13}{mag}$. At closer distances we have too few stars for trustworthy estimates due to the bright magnitude limits of both surveys. Beyond that distance range, individual estimates become too uncertain. We do not expect many stars beyond that distance due to the faint magnitude limits, so we do not report values outside this range. We find that the extinction law varies with each line of sight and along the line of sight, supporting previous works which contend that using a single value to parametrize extinction is insufficient to properly model the three dimensional dust distribution in the Galaxy. The data are available via the \textit{MNRAS} website.

As previously discussed in \citet{Hanson2014}, the key limitation at this stage is the distance inference, which is limited by photometric errors and intrinsic model degeneracies. Furthermore, on the account of our use of stellar models to estimate stellar effective temperatures, there are likely to be systematic uncertainties in our estimates of $A_0$ and $R_0$. These enter through the assumption of 'true' model temperatures, the use of an HRD prior and lack of metallicity variations \citep[again, see][]{Hanson2014}. Furthermore, our extinction estimates for individual lines of sight do not account for correlations in angular dimensions. That is, neighbouring lines-of-sight are solved for independently. This clearly does not mirror reality, where the extinction estimates for stars that are close in space (and whose photons are affected by the same dust structures) should be strongly correlated, whereas those of stars that have a large separation should be less so. Theoretically, due to the finite cross-sectional area of a line-of-sight, a more distant star could show less extinction. This shortcoming is now starting to be addressed. \citet{SaleMagorrian2014} introduce a method based on Gaussian random fields and a model of interstellar turbulence, which addresses the discontinuities we currently see in most extinction maps. \citet{Lallement2014} use an inversion method with spatial correlation kernels that attempts to reconstruct structures of the ISM in a more realistic manner.

Combining current large area photometric surveys, such as those employed here, with parallax measurements from \textit{Gaia} will enable us to construct accurate 3D maps of stars in the Galaxy. Including stellar parameter estimates from future data releases by the Data Processing and Analysis Consortium (\textit{DPAC}), as summarised in \citet{Bailer-Jones2013}, will significantly increase our capabilities of reconstructing the full three dimensional distribution of dust.

\section*{Acknowledgments}
We thank the referees for constructive comments and suggestions. We thank E.~F. Schlafly and H.-W. Rix for helpful discussions. This project is funded by the Sonderforschungsbereich SFB881 'The Milky Way System' (subproject B5) of the German Research Foundation (DFG). RJH is member of the International Max-Planck Research School for Astronomy and Cosmic Physics at the University of Heidelberg (IMPRS-HD) and the Heidelberg Graduate School of Fundamental Physics (HGSFP). 

The Pan-STARRS1 Surveys (PS1) have been made possible through contributions of the Institute for Astronomy, the University of Hawaii, the Pan-STARRS Project Office, the Max-Planck Society and its participating institutes, the Max Planck Institute for Astronomy, Heidelberg and the Max Planck Institute for Extraterrestrial Physics, Garching, The Johns Hopkins University, Durham University, the University of Edinburgh, Queen's University Belfast, the Harvard-Smithsonian Center for Astro- physics, the Las Cumbres Observatory Global Telescope Network Incorporated, the National Central University of Taiwan, the Space Telescope Science Institute, the National Aeronautics and Space Administration under grant No. NNX08AR22G issued through the Planetary Science Division of the NASA Science Mission Directorate, the National Science Foundation under grant No. AST-1238877, the University of Maryland, and Eotvos Lorand University (ELTE).

This work is based in part on observations made with the Spitzer Space Telescope, which is operated by the Jet Propulsion Laboratory, California Institute of Technology under a contract with NASA.

Numerical simulations were performed on the Milky Way supercomputer, which is funded by the Deutsche Forschungsgemeinschaft (DFG) through the Collaborative Research Center (SFB 881) "The Milky Way System" (subproject Z2) and hosted and co-funded by the J\"ulich Supercomputing Center (JSC).

This research made use of the cross-match service provided by CDS, Strasbourg.

\appendix
\section[]{Samples of data products.}
\label{app:data}
In Table~\ref{tab:raw} we show the schema for the full set of individual stars with coordinates, APs and confidence intervals. In Table~\ref{tab:sum} we present the schema of the summary 3D maps as presentend in Figures~\ref{fig:extinctionmap-ar} and \ref{fig:extinctionmap-r}. This includes the centres of the cells and the means and uncertainties of $A_{\rm r}$ and $R_0$ as computed using Equations~\ref{eq:extinctionmap-weightedmean} and \ref{eq:extinctionmap-weights}.

The data files are available via the \textit{MNRAS} website.

\begin{table*}
	\setlength{\tabcolsep}{3pt} 
	\begin{minipage}{\textwidth}
		\caption{A sample of the output catalogue of the individual stellar APs. The \textit{CI} columns show the lower and upper $68\%$ confidence interval limits for the four APs. The lower bound of the $68\%$ confidence interval has $16\%$ of the probability below it, whereas the upper bound has $16\%$ of the probability above it. We include the converted $r_{\rm PS1}$-band extinctions. The three final columns denote flags indicating whether an inferred AP fits the forward model well ($0$) or not ($1$). Only those stars whose flags are all $0$ are used to build the 3D maps (see Table~\ref{tab:sum}). The complete dataset for $19\,885\,031$ stars is split into individual files based on latitude and is available via the \textit{MNRAS} website.}
		\label{tab:raw}
		\centering
		{\scriptsize
 		\begin{tabular}{@{}llcccccccccccccccccc@{}}
			\hline
			$l$/deg & $b$/deg 
			& $A_0$/mag & \multicolumn{2}{c}{${\rm CI}_{A_0}$/mag}
			& $A_{\rm r}$/mag & \multicolumn{2}{c}{${\rm CI}_{A_{\rm r}}$/mag}
			& $R_0$ & \multicolumn{2}{c}{${\rm CI}_{R_0}$}
			& $T_{\rm eff}$/K & \multicolumn{2}{c}{${\rm CI}_{T_{\rm eff}}$/K}
			& $\mu$/mag & \multicolumn{2}{c}{${\rm CI}_{\mu}$/mag} 
			& $F_{A_0}$ & $F_{T_{\rm eff}}$ & $F_{R_0}$   \\
			\hline
33.03003 & -0.84075 & 0.03 & 0.03 & 0.03 & 0.03 & 0.02 & 0.03 & 3.36 & 3.27 & 3.46 & 4289.87 & 4188.03 & 4383.79 & 16.72 & 15.37 & 22.18 & 0 & 0 & 0 \\
75.63519 & -0.83792 & 4.71 & 4.53 & 4.86 & 3.96 & 3.81 & 4.08 & 3.78 & 3.67 & 3.93 & 4650.18 & 4492.93 & 4774.63 & 10.74 & 9.86 & 11.72 & 0 & 0 & 0 \\
190.67752 & -0.83848 & 3.46 & 3.38 & 3.62 & 2.91 & 2.85 & 3.05 & 3.65 & 3.43 & 3.83 & 5633.18 & 5499.33 & 5917.11 & 12.20 & 8.38 & 14.73 & 0 & 0 & 0 \\
61.84478 & -0.83407 & 0.03 & 0.03 & 0.04 & 0.03 & 0.03 & 0.04 & 3.37 & 3.28 & 3.47 & 4179.51 & 4088.93 & 4279.08 & 8.42 & 5.78 & 13.25 & 0 & 0 & 0 \\
228.24329 & -0.83960 & 0.18 & 0.14 & 0.28 & 0.16 & 0.12 & 0.24 & 5.14 & 4.88 & 6.05 & 4263.23 & 4162.75 & 4362.75 & 17.10 & 16.02 & 18.37 & 0 & 0 & 0 \\
161.63760 & -0.84378 & 0.03 & 0.01 & 0.04 & 0.02 & 0.01 & 0.04 & 3.61 & 2.52 & 5.21 & 3885.80 & 3799.36 & 3981.99 & 18.62 & 15.86 & 28.00 & 0 & 0 & 0 \\
195.39756 & -0.83678 & 0.78 & 0.63 & 1.08 & 0.69 & 0.55 & 0.95 & 5.06 & 4.78 & 5.88 & 4200.52 & 4055.83 & 4332.12 & 11.37 & 9.87 & 13.06 & 0 & 0 & 0 \\
182.35284 & -0.83555 & 1.23 & 1.16 & 1.28 & 1.09 & 1.03 & 1.14 & 5.88 & 5.74 & 6.01 & 3890.57 & 3804.96 & 3989.51 & 8.77 & 8.34 & 9.42 & 0 & 0 & 1 \\
18.39156 & -0.84392 & 2.86 & 2.75 & 2.95 & 2.40 & 2.31 & 2.48 & 3.66 & 3.55 & 3.80 & 4627.33 & 4492.73 & 4730.10 & 15.01 & 13.03 & 18.10 & 0 & 0 & 0 \\
166.18062 & -0.83409 & 0.91 & 0.82 & 0.98 & 0.76 & 0.69 & 0.82 & 3.21 & 3.00 & 3.38 & 5710.52 & 5481.76 & 5889.11 & 14.20 & 11.60 & 17.73 & 0 & 0 & 0 \\
			\hline
		\end{tabular}
 	}
	\end{minipage}
\end{table*}

\begin{table*}
	\begin{minipage}{\textwidth}
		\caption{Schema of the summarised 3D map data as presented in Figures~\ref{fig:extinctionmap-ar} and \ref{fig:extinctionmap-r}. The coordinates describe the centres of the cell at a resolution of $\unit{7}{'}$ in both $l$ and $b$. The distance slices have a separation of $\unit{1}{mag}$ in distance modulus. In total there are $322\,207$ cells with data. The full dataset is split into individual files based on the seven slices in distance modulus.}
		\label{tab:sum}
		\centering
		\begin{tabular}{@{}ccccccc@{}}
			\hline
			$l$/deg & $b$/deg & $\mu$/mag 
			& $A_{\rm r}$/mag & $\sigma_{A_{\rm r}}$/mag 
    			& $R_0$ & $\sigma_{R_0}$ \\
    			\hline
			0.40833 & 0.89167 & 8 & 0.96 & 0.38 & 4.97 & 0.34 \\
			0.40833 & 1.00833 & 8 & 1.20 & 0.43 & 4.88 & 0.48 \\
			0.40833 & 1.59167 & 8 & 1.62 & 0.56 & 4.20 & 0.43 \\
			0.40833 & 1.94167 & 8 & 1.74 & 0.63 & 4.10 & 0.54 \\
			0.40833 & 2.05833 & 8 & 2.13 & 0.52 & 3.67 & 0.29 \\
			\hline
		\end{tabular}
	\end{minipage}
\end{table*}

\label{lastpage}


\begin{thebibliography}{99}
\bibitem[\protect\citeauthoryear{Bailer-Jones}{2011}]{Bailer-Jones2011} Bailer-Jones C.~A.~L., 2011, MNRAS, 411, 435
\bibitem[\protect\citeauthoryear{Bailer-Jones et al.}{2013}]{Bailer-Jones2013} Bailer-Jones C.~A.~L. et al., 2013, A\&A, 559, A74
\bibitem[\protect\citeauthoryear{Benjamin et al.}{2003}]{Benjamin2003} Benjamin R.~A. et al., 2003, PASP, 115, 953
\bibitem[\protect\citeauthoryear{Berry et al.}{2012}]{Berry2012} Berry M. et al., 2012, ApJ, 757, 166
\bibitem[\protect\citeauthoryear{Casali et al.}{2007}]{Casali2007} Casali M. et al., 2007, A\&A, 467, 777 
\bibitem[\protect\citeauthoryear{Churchwell et al.}{2009}]{Churchwell2009} Churchwell E. et al., 2009, PASP, 121, 213
\bibitem[\protect\citeauthoryear{Chen et al.}{2013}]{Chen2013} Chen B.-Q., Schultheis M., Jiang B.~W., Gonzalez O.~A., Robin A.~C., Rejkuba M., Minniti, D., 2013, A\&A, 550, A42
\bibitem[\protect\citeauthoryear{Chen et al.}{2014}]{Chen2014} Chen B.-Q. et al., 2014, MNRAS, 443, 1192
\bibitem[\protect\citeauthoryear{Dotter et al.}{2008}]{Dotter2008} Dotter A., Chaboyer B., Jevremovi\'{c} D., Kostov V., Baron E., Ferguson J.~W., 2008, ApJS, 178, 89
\bibitem[\protect\citeauthoryear{Fazio et al.}{2004}]{Fazio2004} Fazio G.~G. et al., 2004, ApJS, 154, 10
\bibitem[\protect\citeauthoryear{Fitzpatrick}{1999}]{Fitzpatrick1999} Fitzpatrick E., 1999, PASP, 111, 63
\bibitem[\protect\citeauthoryear{Goa, Jiang \& Li}{2009}]{Gao2009} Gao J., Jiang B. W., Li A., 2009, ApJ, 707, 89 
\bibitem[\protect\citeauthoryear{Gontcharov}{2012}]{Gontcharov2012} Gontcharov G.~A., 2012, AstL, 38, 12
\bibitem[\protect\citeauthoryear{Gonzalez et al.}{2011}]{Gonzalez2011} Gonzalez O.~A., Rejkuba M., Zoccali M., Valenti E., Minniti D., 2011, A\&A, 534, A3 
\bibitem[\protect\citeauthoryear{Gonzalez et al.}{2012}]{Gonzalez2012} Gonzalez O.~A., Rejkuba M., Zoccali M., Valenti E., Minniti D., Schultheis M., Tobar R., Chen, B., 2012, A\&A, 543, A13 
\bibitem[\protect\citeauthoryear{Green et al.}{2014}]{Green2014} Green G.~M. et al., 2014, ApJ, 783, 114
\bibitem[\protect\citeauthoryear{Gustafsson et al.}{2008}]{Gustafsson2008} Gustafsson B., Edvardsson B., Eriksson K., J\o rgensen U.~G., Nordlund \r{A}., Plez, B., 2008, A\&A, 486, 951
\bibitem[\protect\citeauthoryear{Hanson \& Bailer-Jones}{2014}]{Hanson2014} Hanson R.~J., Bailer-Jones C.~A.~L., 2014, MNRAS, 438, 2938
\bibitem[\protect\citeauthoryear{Jones, West \& Foster}{2011}]{Jones2011} Jones D.~O., West A.~A., Foster J.~B., 2011, AJ, 142, 44
\bibitem[\protect\citeauthoryear{Kaiser et al.}{2010}]{Kaiser2010} Kaiser N. et al., 2010, Proc. SPIE, 7733, 77330
\bibitem[\protect\citeauthoryear{Lallement et al.}{2014}]{Lallement2014} Lallement R., Vergely J.-L., Valette B., Puspitarini L., Eyer L., Casagrande, L., 2014, A\&A, 561, 91
\bibitem[\protect\citeauthoryear{Marshall et al.}{2006}]{Marshall2006} Marshall D.~J., Robin A.~C., Reyl\'{e} C., Schultheis M., Picaud, S., 2006, A\&A, 453, 635
\bibitem[\protect\citeauthoryear{M\"ortsell}{2013}]{Moertsell2013} M\"ortsell E., 2013, A\&A, 550, A80
\bibitem[\protect\citeauthoryear{Sale}{2012}]{Sale2012} Sale S.~E., 2012, MNRAS, 427, 2119
\bibitem[\protect\citeauthoryear{Sale \& Magorrian}{2014}]{SaleMagorrian2014} Sale S.~E., Magorrian J., 2014, MNRAS, 445, 256
\bibitem[\protect\citeauthoryear{Sale et al.}{2014}]{Sale2014} Sale S.~E. et al., 2014, MNRAS, 443, 2907
\bibitem[\protect\citeauthoryear{Savage \& Mathis}{1979}]{Savage1979} Savage B.~D., Mathis J.~S., 1979, ARA\&A, 17, 73
\bibitem[\protect\citeauthoryear{Schlafly et al.}{2012}]{Schlafly2012} Schlafly E.~F. et al., 2012, ApJ, 756, 158
\bibitem[\protect\citeauthoryear{Schlafly et al.}{2014a}]{Schlafly2014a} Schlafly E.~F. et al., 2014, ApJ, 786, 29
\bibitem[\protect\citeauthoryear{Schlafly et al.}{2014b}]{Schlafly2014b} Schlafly E.~F. et al., 2014, ApJ, 789, 15
\bibitem[\protect\citeauthoryear{Schlegel, Finkbeiner \& Davis}{1998}]{Schlegel1998} Schlegel D., Finkbeiner D.~P., Davis M., 1998, ApJ, 500, 525
\bibitem[\protect\citeauthoryear{Skrutskie et al.}{2006}]{Skrutskie2006} Skrutskie M.~F. et al., 2006, AJ, 131, 1163 
\bibitem[\protect\citeauthoryear{Stubbs et al.}{2010}]{Stubbs2010} Stubbs C.~W. Doherty P., Cramer C., Narayan G., Brown Y.~J., Lykke K.~R., Woodward J.~T., Tonry J.~L., 2010, ApJS, 191, 376 
\bibitem[\protect\citeauthoryear{Tonry et al.}{2012}]{Tonry2012} Tonry J.~L. et al., 2012, ApJ, 750, 99
\bibitem[\protect\citeauthoryear{Vergeley et al.}{2010}]{Vergely2010} Vergely J.-L., Valette B., Lallement R., Raimond S., 2010, A\&A, 518, 31
\bibitem[\protect\citeauthoryear{Zasowski et al.}{2009}]{Zasowski2009} Zasowski G. et al., 2009, ApJ, 707, 510

\end{thebibliography}
\end{document}